\documentclass[MSNbibl,nameyear,seceqn,dvips]{arxstspdf}
\usepackage{mathbh}
\usepackage{shere}
\usepackage{graphicx}
\usepackage{flushend}
\usepackage{stfloats}
\volume{28}
\issue{4}
\pubyear{2013}
\firstpage{616}
\lastpage{640}
\doi{10.1214/13-STS443} 

\makeatletter
\fnbelowfloat

\setattribute{abstract}    {skip} {20\p@}
\setattribute{keyword}     {skip} {8\p@}

\setattribute{frontmatter} {cmd}  {}
\setattribute{abstract}   {width}  {345pt}
\setattribute{keyword}    {width}  {345pt}
\newcommand{\cal}{\mathcal}

\newcommand{\citecs}[1]{\citeauthor{#1}, \citeyear{#1}}
\renewcommand{\citep}[1]{(\citeauthor{#1}, \citeyear{#1})}

\newtheorem{theorem}{Theorem}[section]

\newcommand{\real}{{\mathbb{R}}}
\newcommand{\sphere}{{\mathbb{S}}}

\makeatother

\begin{document}
\begin{frontmatter}

\title{Uncertainty Quantification in Complex Simulation
Models Using Ensemble Copula~Coupling}
\runtitle{Ensemble Copula Coupling}
\pdftitle{Uncertainty Quantification in Complex Simulation Models Using Ensemble Copula Coupling}

\begin{aug}
\author[a]{\fnms{Roman} \snm{Schefzik}\ead[label=e1,text=r.schefzik@uni-heidelberg.de]{r.schefzik@uni-heidelberg.de}},
\author[b]{\fnms{Thordis L.} \snm{Thorarinsdottir}\ead[label=e2,text=thordis@nr.no]{thordis@nr.no}}
\and
\author[a]{\fnms{Tilmann} \snm{Gneiting}\corref{}\ead[label=e3,text=t.gneiting@uni-heidelberg.de]{t.gneiting@uni-heidelberg.de}}
\runauthor{R. Schefzik, T. L. Thorarinsdottir and T. Gneiting}

\affiliation{Heidelberg University, Norwegian Computing Center and
Heidelberg University}

\address[a]{Roman Schefzik is Ph.D. Student and Tilmann Gneiting is
Professor, Institute for Applied Mathematics,
Heidelberg University, Im Neuenheimer Feld 294, 69120 Heidelberg, Germany (e-mail: \printead*{e1}; \printead*{e3}).}
\address[b]{Thordis L.~Thorarinsdottir is Senior Research Scientist,
Norwegian Computing Center, P.O. Box 114, Blindern, 0314 Oslo, Norway
\printead{e2}.}

\end{aug}

%
\begin{abstract}
Critical decisions frequently rely on high-dimensional output from
complex computer simulation models that show intricate cross-variable,
spatial and temporal dependence structures, with weather and climate
predictions being key examples. There is a strongly increasing
recognition of the need for uncertainty quantification in such
settings, for which we propose and review a general multi-stage
procedure called ensemble copula coupling (ECC), proceeding as follows:

1. Generate a raw ensemble, consisting of multiple runs of the computer
model that differ in the inputs or model parameters in suitable
ways.

2. Apply statistical postprocessing techniques, such as Bayesian model
averaging or nonhomogeneous regression, to correct for systematic
errors in the raw ensemble, to obtain calibrated and sharp
predictive distributions for each univariate output variable
individually.

3. Draw a sample from each postprocessed predictive distribution.

4. Rearrange the sampled values in the rank order structure of the raw
ensemble to obtain the ECC postprocessed ensemble.

The use of ensembles and statistical postprocessing have become routine
in weather forecasting over the past decade. We show that seemingly
unrelated, recent advances can be interpreted, fused and consolidated
within the framework of ECC, the common thread being the adoption of
the empirical copula of the raw ensemble. Depending on the use of
Quantiles, Random draws or Transformations at the sampling stage, we
distinguish the ECC-Q, ECC-R and ECC-T variants, respectively. We also
describe relations to the Schaake shuffle and extant copula-based
techniques. In a case study, the ECC approach is applied to
predictions of temperature, pressure, precipitation and wind over
Germany, based on the 50-member European Centre for Medium-Range
Weather Forecasts (ECMWF) ensemble.
\end{abstract}

%
\begin{keyword}
\kwd{Bayesian model averaging}
\kwd{empirical copula}
\kwd{ensemble calibration}
\kwd{nonhomogeneous regression}
\kwd{numerical weather prediction}
\kwd{probabilistic forecast}
\kwd{Schaake shuffle}
\kwd{Sklar's theorem}
\end{keyword}

\pdfkeywords{Bayesian model averaging,
empirical copula, ensemble calibration,
nonhomogeneous regression, numerical weather prediction,
probabilistic forecast, Schaake shuffle,
Sklar's theorem}

\end{frontmatter}

\section{Introduction} \label{secintroduction}

In a vast range of applications, critical decisions depend on the
output of complex computer simulation models, with examples including
weather and climate predictions and the management of floods,
wildfires, air quality and groundwater contaminations. There is a
much increased recognition of the need for quantifying the uncertainty
in the model output, as evidenced by the creation of pertinent
American Statistical Association (ASA) and Society for Industrial and
Applied Mathematics (SIAM) interest groups, and by the recent launch
of the SIAM/ASA Journal on Uncertainty Quantification. As SIAM
President Nick \citet{Trefethen2012} notes succinctly,

\begin{quote}
``An answer that used to be a single number may now be a statistical
distribution.''
\end{quote}

\noindent Frequently, the goal is prediction, and we are witnessing a
transdisciplinary change of paradigms in the transition from
deterministic or point forecast to probabilistic or distributional
forecasts \citep{Gneiting2008a}. The goal is to obtain calibrated and
sharp, joint predictive distributions of future quantities of
interest, from which any desired functionals, such as event
probabilities, moments, quantiles and prediction intervals can be
extracted, for a full quantification of the predictive uncertainty.
In this context, calibration refers to the statistical compatibility
of the probabilistic forecasts and the observations, in that events
predicted to occur with probability $p$ ought to realize with
empirical frequency~$p$. Sharpness refers to the concentration of the
predictive distributions and is a property of the probabilistic
forecasts only \citep{GneitingBalabdaouiRaftery2007}. While our data
examples all concern weather forecasting, where the recognition of the
need for uncertainty quantification can be traced at least to
\citet{Cooke1906}, the methods and principles we discuss apply in much
broader contexts, both predictive and in other settings, where one
seeks to quantify the uncertainty in our incomplete knowledge of
current or past quantities and events.

%
%
\begin{figure*}

\includegraphics{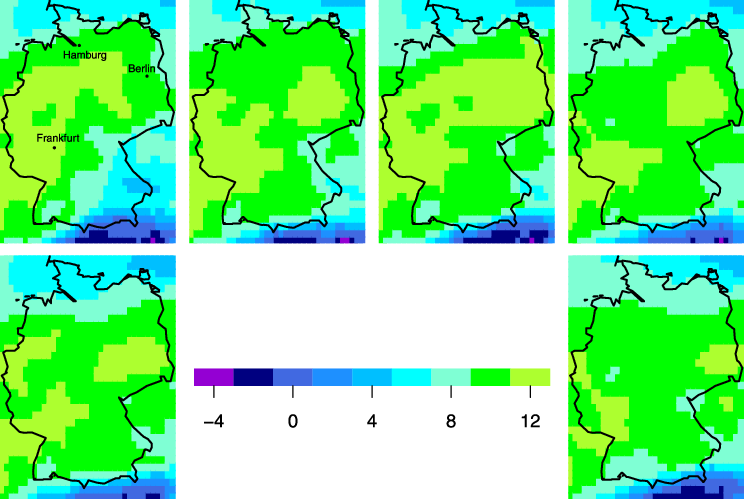}

\caption{48-hour ahead ECWMF ensemble forecast for temperature over
Germany valid 2:00 am on April 1, 2011, in the unit of degrees
Celsius. Six randomly selected members are shown. The top left
panel shows the locations of the three stations used in the
subsequent case study.}
\label{figenstemp}
\end{figure*}
%
\begin{figure*}[b]

\includegraphics{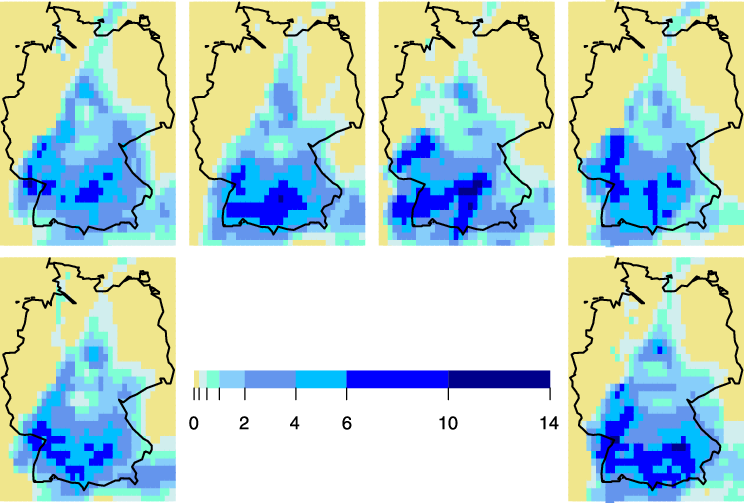}

\caption{24-hour ahead ECWMF ensemble forecast for six-hour
precipitation accumulation over Germany valid 2:00 am on May 20,
2010, in the unit of millimeters. Six randomly selected
members are shown.} \label{figensprecip}
\end{figure*}
%

Focusing attention on the setting of our case study, accurate
predictions of future weather are of considerable value for society.
Medium-range weather forecasts, with lead times up to two weeks, are
obtained by numerically solving the partial differential equations
that describe the physics of the atmosphere, with initial conditions
provided by estimates of the current state of the atmosphere\vadjust{\goodbreak}
\citep{Kalnay2003}. In order to account for the uncertainties in the
forecast, national and international meteorological centers use
ensembles of numerical weather prediction (NWP) model output, where
the ensemble members differ in terms of the two major sources of
uncertainty, namely, the initial conditions and the parameterization
of the NWP model (\citecs{Palmer2002}; \citecs{GneitingRaftery2005}).
To give an
example, Figures~\ref{figenstemp} and \ref{figensprecip}
illustrate forecasts of surface temperature and six-hour precipitation
accumulation\break  over Germany issued by the European Centre for
Medium-Range Weather Forecasts (ECMWF) as a part of its 50-member
real-time ensemble, which operates at a horizontal resolution of
approximately 32 km and lead times up to ten days (\citecs{Molteni1996};
\citecs{LeutbecherPalmer2008}). The valid time of these forecasts is 00:00
Universal Time Coordinated (UTC) in meteorological format, which we
convert to local time in what follows.

While the goal of NWP ensemble systems is to capture the inherent
uncertainty in the prediction, they are subject to systematic errors,
such as biases and dispersion errors. It is therefore common practice
to statistically postprocess the output of NWP ensemble forecasts,
with state of the art techniques including the ensemble Bayesian model
averaging (BMA) approach developed by \citet{Raftery2005} and the
nonhomogeneous regression (NR) or ensemble model output statistics
(EMOS) technique proposed by \citet{Gneiting2005}.

\begin{figure*}

\includegraphics{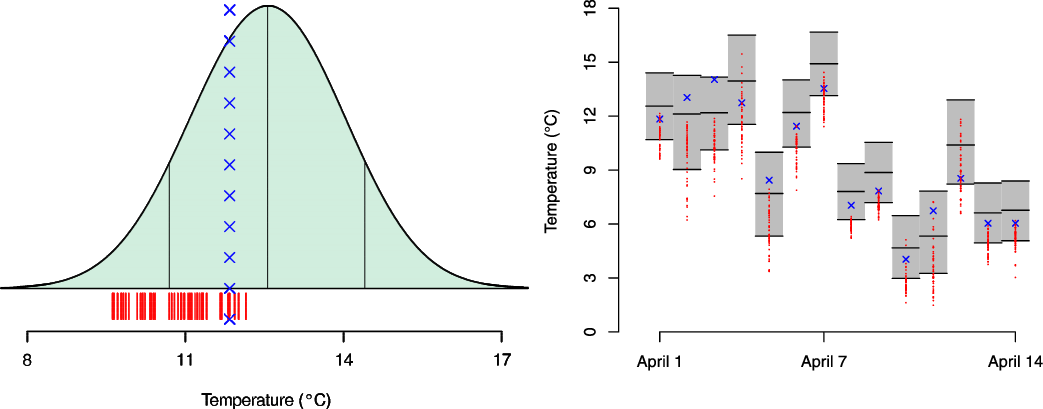}

\caption{48-hour ahead BMA postprocessed predictive distributions for
temperature in Berlin based on the 50-member ECMWF ensemble. The
ensemble forecast is shown in red, the realizing observation in
blue. Left: predictive density valid 2:00 am on April 1, 2011.
Right: 10th, 50th and 90th percentiles of the predictive
distributions valid 2:00 am on April 1--14, 2011.}
\label{figpredPDFtemp}\vspace*{3pt}
\end{figure*}

\begin{figure*}[b]\vspace*{6pt}

\includegraphics{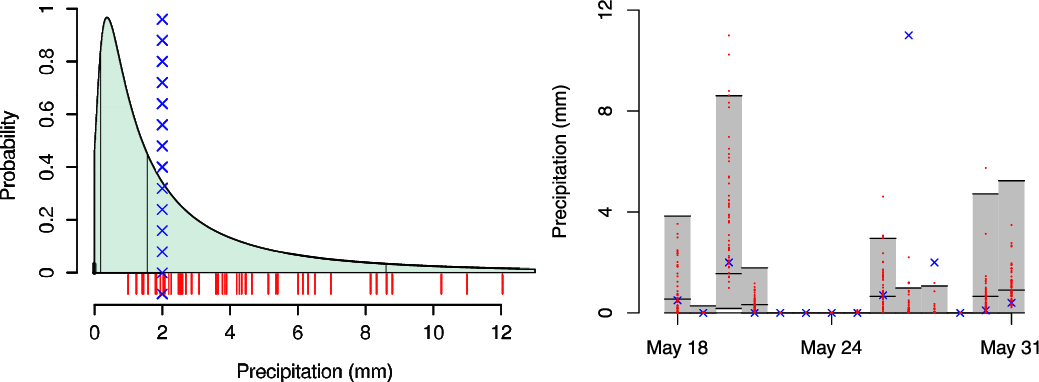}

\caption{24-hour ahead BMA postprocessed predictive distributions for
six-hour precipitation accumulation in Frankfurt based on the
50-member ECMWF ensemble. The ensemble forecast is shown in red,
the realizing observation in blue. Left: mixed discrete-continuous
predictive distribution valid 2:00 am on May 20, 2010, comprising a
point mass of 0.033 at zero, which is indicated by the thick black
bar, and a density at positive accumulations, with mass 0.967.
Right: 10th, 50th and 90th percentiles of the predictive
distribution distributions valid 2:00 am on May 18--31, 2010.}
\label{figpredPDFprecip}
\end{figure*}

To illustrate the idea, let $y$ denote the weather quantity of
interest, such
as temperature at a specific location and look-ahead time, and write
$x_1,\ldots,\break  x_M$ for the corresponding $M$ ensemble member
forecasts. The ensemble BMA approach employs mixture distributions of
the general form
\[
y | x_1,\ldots, x_M \sim\sum
_{m=1}^M w_m f(y | x_m),
\]
where the left-hand side refers to the conditional distribution given
the ensemble member forecasts. Here $f(y | x_m)$ denotes a
parametric probability distribution or kernel that depends on the
ensemble member forecast $x_m$ in suitable ways, with the mixture
weights $w_1,\ldots, w_M$ reflecting the members' relative
contributions to predictive skill over a training period. BMA
postprocessed predictive distributions based on the 50-member ECMWF
ensemble are illustrated in Figure~\ref{figpredPDFtemp} for
temperature, where the kernel is normal and the postprocessing
corrects for both a low bias and underdispersion, and in Figure~\ref{figpredPDFprecip} for precipitation, where the kernel
comprises a point mass at zero along with a power transformed gamma
distribution for positive accumulations.

%
%
%

%
%
%

In contrast, the NR predictive distribution is a single
parametric distribution of the general form
\[
y | x_1,\ldots, x_M \sim g(y | x_1,\ldots,
x_M),
\]
where $g$ is a parametric distribution function with location, scale
and shape parameters depending on the ensemble values in suitable
ways. For example, $g$ could be normal with the mean an affine function
of the ensemble member forecasts and the variance an affine function of
the ensemble variance.

Statistical postprocessing techniques such as ensemble BMA and NR have
been shown to substantially improve the predictive skill of the NWP
ensemble output (\citecs{WilksHamill2007}; Hagedorn et~al., \citeyear{Hagedorn2012}). Frequently,
such methods apply to each weather variable at each location and each
lead time individually and, therefore, they may fail to take
cross-variable, spatial and temporal interactions properly into
account. NWP models rely on discretizations of the equations that
govern the physics of the atmosphere and, thus, multivariate dependence
structures tend to be reasonably well represented in the raw ensemble
system. However, these structures may fail to be retained if the
univariate margins are postprocessed individually. In low-dimensional
or highly structured settings, parametric approaches to the modeling
of multivariate dependence structures in the forecast errors are
feasible, such as in the recent work of \citet{Pinson2011},
\citet{Schuhen2012} and \citet{Sloughter2011} on wind vectors, or in
the approach of \citet{Gel2004} and \citet{Berrocal2007} that relies
on geostatistical models in spatial settings.

However, the statistical postprocessing of a full NWP ensemble
forecast poses extremely high-dimen\-sional problems. For instance, we
might be interested in five weather variables at $500 \times500$ grid
boxes, ten vertical levels and 72 lead times, for a total of 900
million variables. While not all of them may need to be considered
simultaneously, critical applications, such as air traffic control
(Chaloulos and Lygeros, \citeyear{ChaloulosLygeros2007}), air quality \citep{DelleMonache2006}
and flood management (Cloke and
Pappenberger, \citeyear{ClokePappenberger2009}; \citecs{Schaake2010}),
depend on physically realistic probabilistic forecasts of
spatio-tempo\-ral weather trajectories and therefore may entail much
higher dimensions than can readily be incorporated into a parametric
model.

%
%
\begin{figure*}[t]
\begin{tabular}{@{}c@{\hspace*{4pt}}c@{}}

\includegraphics{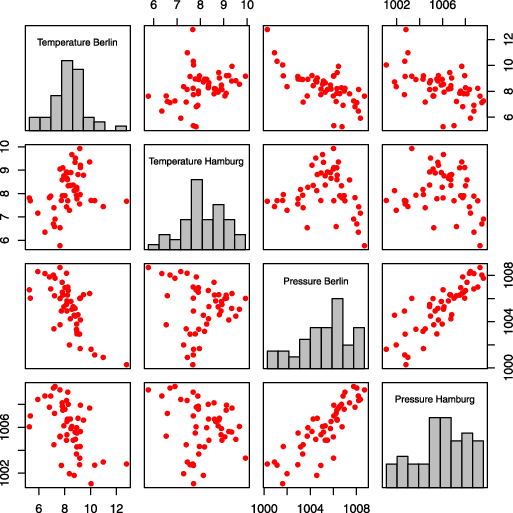}
 & \includegraphics{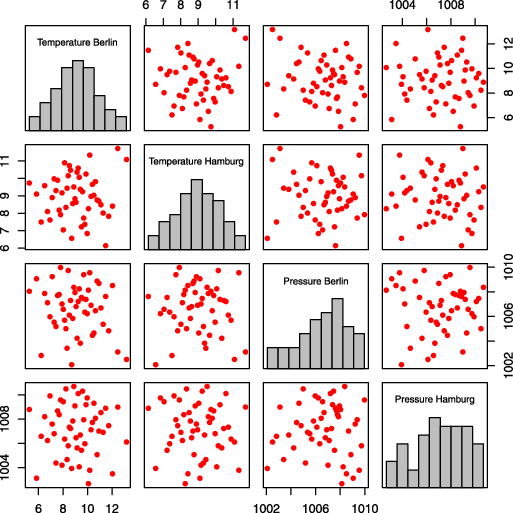}\\
(a) Raw ECMWF ensemble & (b) Individual BMA postprocessing\\[8pt]
\multicolumn{2}{@{}c@{}}{
\includegraphics{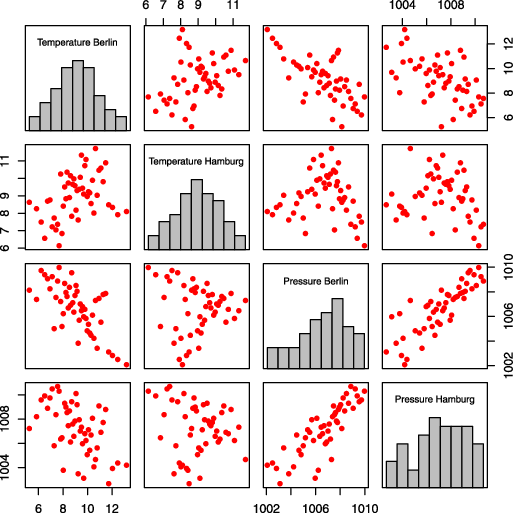}
}\\
\multicolumn{2}{@{}c@{}}{(c) ECC postprocessed ensemble}
\end{tabular}
\caption{24-hour ahead ensemble forecasts of temperature and pressure
at Berlin and Hamburg, valid 2:00 am on May 27, 2010. The units
used are degrees Celsius and hPa.} \label{figECC}\vspace*{-6pt}
\end{figure*}
%

To address this challenge, we propose and review a general multi-stage
procedure called ensemble copula coupling (ECC), originally hinted at
by \citet{Bremnes2007} and \citet{KrzysztofowiczToth2008}, and recently
investigated and developed by \citet{Schefzik2011}. The ECC approach
allows for the multivariate rank dependence structure of the raw NWP
ensemble to be preserved in the postprocessed ensemble, proceeding
roughly as follows.

\begin{longlist}
\item[\textit{Univariate postprocessing}.] Apply statistical postprocessing
techniques, such as ensemble BMA or NR, to obtain calibrated and
sharp marginal predictive distributions for each weather variable,
location and look-ahead time individually.

\item[\textit{Quantization}.] Draw a discrete sample of the same size as the raw
ensemble from each univariate, postprocessed predictive
distribution.

\item[\textit{Ensemble reordering}.] Arrange the sampled values in the rank
order structure of the raw ensemble to obtain the ECC
postprocessed ensemble.
\end{longlist}

An illustration of the ECC approach is given in Figure~\ref{figECC},
a dynamic version of which is available in the supplementary\vadjust{\goodbreak} material
(Schefzik, Thorarinsdottir and Gneiting, \citeyear{Schefzik2013}). Here, the setting is four dimensional. We
consider surface temperature and sea level pressure in Berlin and
Hamburg, respectively. The scatterplot matrix in the top panel
illustrates the 50-member ECMWF ensemble forecast at a 24 hours lead
time. Clearly, there are dependencies between the margins; for
example, there is a positive association between temperature in Berlin
and temperature in Hamburg, and there are negative associations
between temperature and pressure. The scatterplot matrix in the
middle panel is constructed from samples of the individually BMA
postprocessed predictive distributions. Here, the systematic errors
in the margins have been corrected, at the cost of a loss of the error
dependence structure. The bottom panel elucidates the effects of the
ECC ensemble reordering; while the margins remain unchanged from the
middle panel, the rank dependence structure of the raw ensemble is
restored.

Owing to the intuitive appeal and striking simplicity, which incurs
essentially no computational costs beyond the marginal postprocessing,
approaches of ECC type are rapidly gaining prominence at weather
centers worldwide, with variants recently having been implemented by
\citet{Flowerdew2012}, \citet{Pinson2011} and
\citet{RoulinVannitsem2012}, among others. Our goal here is to
interpret, fuse and consolidate these and other seemingly unrelated
advances within the framework of ECC. As we will demonstrate, the
common thread of the approaches lies in the adoption of the empirical
copula of the raw ensemble, thereby restoring its rank dependence
structure and justifying the term ensemble copula coupling.

The remainder of the paper is organized as follows. In Section~\ref{secunivariate} we review and discuss statistical postprocessing
techniques for univariate NWP ensemble output. General copula
approaches to the handling of multivariate output are discussed in
Section~\ref{seccopulas}, with subsequent focus on the ECC approach
in Section~\ref{secECC}, where we distinguish the ECC-Q, ECC-R and
ECC-T variants, depending on the use of Quantiles, Random draws or
Transformations at the quantization stage. Section~\ref{secdata}
turns to a case study on probabilistic predictions of temperature,
pressure, precipitation and wind over Germany, based on the ECMWF
ensemble. The paper closes with Section~\ref{secdiscussion}, where we
discuss benefits and limitations of the ECC approach and return to the
general theme of uncertainty quantification for high-dimensional
output from complex simulation models with intricate dependence
structures.

%
%
\begin{table*}
\tablewidth=320pt
\caption{Ensemble BMA implementations for univariate weather
quantities. In the case of precipitation amount, we refer to
$y^{1/3} \in\real^+$, because the gamma kernels apply to cube root
transformed precipitation accumulations. In the case of wind
direction, $\sphere$ denotes the circle, $z_m$ is a bias-corrected
ensemble member value on the circle, and $\kappa_m$ is a
concentration parameter, for $m = 1,\ldots, M$}
\label{tabBMA}
\begin{tabular*}{\tablewidth}{@{\extracolsep{\fill}}lcccc@{}}
\hline
\textbf{Weather quantity} & \textbf{Range} & \textbf{Kernel ($\bolds{f}$)} & \textbf{Mean}
& \textbf{Variance} \\
\hline
Temperature & $y \in\real$ & Normal & $a_m + b_m x_m$ & $\sigma_m^2$ \\
Pressure & $y \in\real$ & Normal & $a_m + b_m x_m$ & $\sigma_m^2$ \\
Precipitation amount
& $y^{1/3} \in\real^+$ & Gamma & $a_m + b_m x_m^{1/3}$ & $c_m + d_m
x_m$ \\
Wind speed
& $y \in\real^+$ & Gamma & $a_m + b_m x_m$ & $c_m + d_m x_m$
\\
Wind direction & $y \in\sphere$ & von Mises & $z_m$ & $\kappa_m^{-1}$
\\
Visibility
& $y \in[0,1]$ & Beta & $a_m + b_m x_m^{1/2}$ & $c_m + d_m x_m^{1/2}$
\\
\hline
\end{tabular*}  \vspace*{-6pt}
\end{table*}

\section{Univariate Postprocessing: Bayesian Model Averaging (BMA) and
Nonhomogeneous Regression (NR)}
\label{secunivariate}

Following the pioneering work of Hamill and\break  Colucci
(\citeyear{HamillColucci1997}), various types of statistical
postprocessing techniques for the output of NWP ensemble forecasts have
been developed, with \citet{WilksHamill2007},
\citet{BroeckerSmith2008}, \citet{SchmeitsKok2010} and
\citet{RuizSaulo2012} providing critical reviews. As noted,
postprocessing aims to correct for biases and dispersion errors in the
ensemble output, and state-of-the-art techniques can roughly be divided
into mixture approaches, building on the ensemble Bayesian model
averaging\break  (BMA) approach of \citet{Raftery2005}, and regression
approaches, such as the nonhomogeneous regression (NR) method put forth
by \citet{Gneiting2005}.\looseness=-1

Specifically, consider a univariate weather quantity of interest, $y$,
and write $x_1,\ldots, x_M$ for the corresponding $M$ ensemble member
forecasts. As noted, the ensemble BMA approach uses mixture
distributions of the general form
%
%
\begin{equation}
\label{eqBMA} y | x_1,\ldots, x_M \sim\sum
_{m=1}^M w_m f(y | x_m),
\end{equation}
where the left-hand side refers to the conditional distribution of $y$
given the ensemble member forecasts $x_1,\ldots, x_M$, and $f(y |
x_m)$ is a parametric distribution that depends on $x_m$
only.\footnote{In the case of ensembles with nonexchangeable members
the distribution $f$ might depend on member specific statistical
parameters. Furthermore, in some implementations $f$ might depend on
observed variables or on NWP model output for quantities other than
$y$, such as in the approach of \citet{Glahn2009}. Similar comments
apply to the NR technique.} The mixture weights $w_1,\ldots, w_m$
are nonnegative and sum to 1; they reflect the corresponding member's
relative contributions to predictive skill over a training period. In
contrast, the NR predictive distribution is a single parametric
distribution of the general form
%
%
\begin{equation}
\label{eqNR} y | x_1,\ldots, x_M \sim g(y |
x_1,\ldots, x_M),
\end{equation}
where the right-hand side refers to a parametric family
of probability distributions, with the parameters depending on all
ensemble members simultaneously.

The particular choice of a parametric model for the BMA kernel $f$
or\vadjust{\goodbreak}
the NR distribution $g$ depends on the weather quantity at hand.
Table~\ref{tabBMA} sketches ensemble BMA implementations for
temperature and pressure \citep{Raftery2005}, where the kernel $f(y
| x_m)$ is normal with mean $a_{0m} + a_{1m} x_m$ and variance
$\sigma_m^2$, precipitation (\citeauthor{Sloughter2007},\break  \citeyear{Sloughter2007}),
wind speed
\citep{SloughterGneitingRaftery2010}, wind direction \citep{Bao2010}
and visibility \citep{ChmieleckiRaftery2010}. Furthermore, ensemble
BMA implementations are available for fog \citep{RoquelaureBergot2008},
visibility and ceiling \citep{ChmieleckiRaftery2010}. Frequently, the
parameters in the specifications for the mean and the variance of the
kernels are subject to constraints; for example, the variance parameters
are often assumed to be constant across ensemble members. If the
ensemble is generated in such a way that its members are statistically
indistinguishable or exchangeable, as in the case of the ECMWF
ensemble, the BMA weights as well as the BMA mean and variance
parameters are assumed to be constant across ensemble members
(\citeauthor{FraleyRafteryGneiting2010},\break  \citeyear{FraleyRafteryGneiting2010}).
Table~\ref{tabNR} hints at NR
implementations for temperature and pressure \citep{Gneiting2005},
where the postprocessed predictive distribution is normal with mean $a
+ b_1 x_1 + \cdots+ b_M x_M$ and variance $c + d S^2$ where $S^2$ is
the ensemble variance, for precipitation (\citecs{Wilks2009}; \citecs
{Scheuerer2013})
and for wind speed (\citecs{ThorarinsdottirGneiting2010};
Thora\-rinsdottir and Johnson, \citeyear{ThorarinsdottirJohnson2011}).

%
\begin{table}
\caption{NR implementations for univariate weather quantities. In the
case of precipitation amount, we refer to the distinct approaches of
Wilks (\citeyear{Wilks2009}) and Scheuerer (\citeyear{Scheuerer2013})}
\label{tabNR}
\begin{tabular*}{\tablewidth}{@{\extracolsep{\fill}}lcc@{}}
\hline
\textbf{Weather quantity} & \textbf{Range} & \textbf{Distribution ($\bolds{g}$)} \\
\hline
Temperature & $y \in\real$ & Normal \\
Pressure & $y \in\real$ & Normal \\
Precipitation amount & $y \in\real^+$ & Truncated logistic \\
& $y \in\real^+$ & Generalized extreme value \\
Wind components & $y \in\real$ & Normal \\
Wind speed & $y \in\real^+$ & Truncated normal \\
\hline
\end{tabular*}
\end{table}

In the remainder of this section we provide a detailed description of
the postprocessing methods for the weather variables temperature,
pressure, precipitation and wind which are analyzed in our case study.
Generally, the ensemble BMA method is more flexible, while the NR
technique is more parsimonious. In terms of the predictive
performance, the general\vadjust{\goodbreak} experience is that the BMA and NR approaches
yield comparable results. Software for estimation and prediction is
available in the form of the \texttt{ensembleBMA} \citep{Fraley2011} and
\texttt{ensembleMOS} packages in \texttt{R}.\footnote{These packages are
available for download at \href{http://www.r-project.org}{www.r-}
\href{http://www.r-project.org}{project.org}.}

\subsection{Temperature and Pressure} \label{sectemperature}

For the weather variables temperature and pressure,
\citet{Raftery2005} propose the ensemble BMA specification
%
%
\begin{equation}
\label{eqBMAtemp}\qquad y | x_1,\ldots, x_M \sim\sum
_{m=1}^M w_m {\cal N}
\bigl(a_m + b_m x_m, \sigma_m^2
\bigr),
\end{equation}
where ${\cal N}( \mu, \sigma^2)$ denotes a normal distribution with
mean $\mu$ and variance $\sigma^2$. The BMA weights $w_1,\ldots,\allowbreak w_M$,
the mean parameters $a_1,\ldots, a_M$ and $b_1,\ldots, b_M$, and the
variance parameters $\sigma_1^2,\ldots, \sigma_M^2$, which in the
standard implementation are assumed to be constant across ensemble
members, are estimated on training data. This type of mixture approach
has been applied successfully at weather centers worldwide,\footnote{A
real-time ensemble BMA implementation for predictions of temperature
and precipitation over the Pacific Northwest region of the United
States is available to the general public at
\href{http://www.probcast.com}{www.probcast.com}, based on the
University of Washington mesoscale ensemble in the form described by
\citet{EckelMass2005}.} and we give an example in
Figure~\ref{figpredPDFtemp}.

\citet{Gneiting2005} propose an NR approach for temperature and pressure,
in which the predictive distribution is normal,
%
%
\begin{eqnarray}
\label{eqNRtemp}\quad
&&y | x_1,\ldots, x_M\nonumber\\[-8pt]\\[-8pt]
&&\quad \sim{\cal{N}}
\bigl(a + b_1 x_1 + \cdots+ b_M
x_M, c + d S^2\bigr),\nonumber
\end{eqnarray}
where $S^2 = \sum_{m=1}^M (x_m - \bar{x})^2/M$ denotes the ensemble
variance. If the ensemble members are exchangeable, it needs to be
assumed that $b_1 = \cdots= b_M$. This approach has also  been
applied at weather centers internationally, as exemplified in the work
of \citet{Hagedorn2008} and \citet{Kann2009}.

\subsection{Precipitation} \label{secprecip}

While of critical applied importance, probabilistic forecasts for
quantitative precipitation pose technical challenges, in that the
predictive distribution is mixed discrete-continuous, comprising both
a point mass at zero and a density on the positive real axis, which
might be considerably skewed.

\citet{Sloughter2007} propose an ensemble BMA model of the general
form (\ref{eqBMA}) for precipitation accumulation, where the kernel
$f(y | x_m)$ is a Bernoulli--Gamma mixture. The Bernoulli
component provides a point mass at zero via a logistic regression
link, in that
%
%
\begin{eqnarray}
\label{eqBMAprecip0}\quad \operatorname{logit}f[y=0 | x_m] &=& \log
\frac{f[y=0 | x_m]}{f[y>0 | x_m]} \nonumber\\[-8pt]\\[-8pt]
&=& \alpha_m + \beta_m
x_m^{1/3} + \gamma_m \delta_m,\nonumber
\end{eqnarray}
where $\delta_m$ equals 1 if $x_m = 0$ and equals 0 otherwise. The
continuous part of the kernel is a gamma distribution in terms of the
cube root transformation, $y^{1/3}$, of the precipitation
accumulation, so that
%
%
\begin{eqnarray}
\label{eqBMAprecip1} f\bigl(y^{1/3} | x_m\bigr) &=& f[y=0 |
x_m] \mathbh{1}_{\{y=0\}} \nonumber\\[-8pt]\\[-8pt]
&&{}+ f[y>0 | x_m] h
\bigl(y^{1/3} | x_{m}\bigr) \mathbh{1}_{\{y>0\}},\hspace*{-29pt}\nonumber
\end{eqnarray}
where $h$ denotes a gamma distribution with mean $\mu_m$ and variance
$\sigma_m^2$, with
%
%
\begin{equation}
\label{eqBMAprecip2}\qquad \mu_m = a_m + b_m
x_m^{1/3} \quad\mbox{and}\quad \sigma_m^2 =
c_m + d_m x_m,
\end{equation}
and where $\mathbh{1}_A$ denotes the indicator function of the event
$A$. Figure~\ref{figpredPDFprecip} shows an example of the
resulting BMA postprocessed predictive distribution in terms of the
nontransformed precipitation accumulation,~$y$.

%
%
\begin{figure*}

\includegraphics{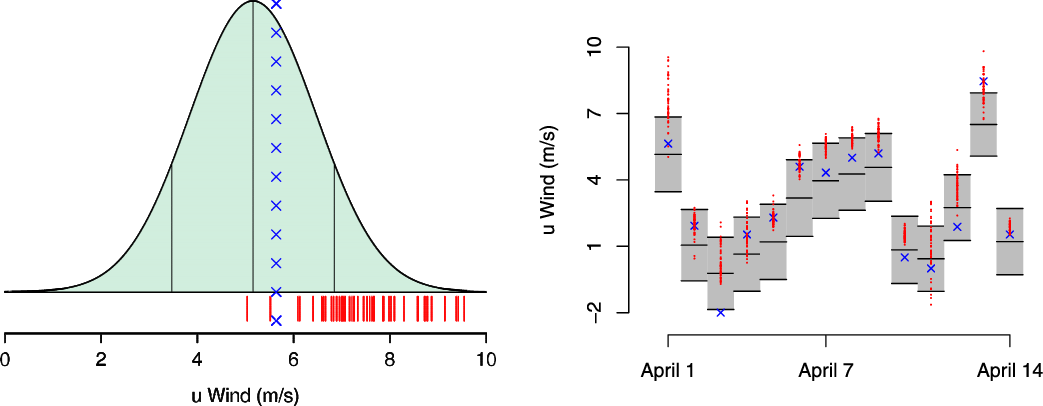}

\caption{24-hour ahead NR postprocessed predictive distributions for
the $u$ wind component at Hamburg based on the 50-member ECMWF
ensemble. The ensemble forecast is shown in red, the realizing
observation in blue. Left: predictive density valid 2:00 am on
April 1, 2011. Right: 10th, 50th and 90th percentiles of the
predictive distributions valid 2:00 am on April 1--14, 2011.}
\label{figpredPDFwind}\vspace*{-3pt}
\end{figure*}
%

Turning to the NR approach, we follow \citet{RoulinVannitsem2012} and
interpret the logistic regression technique of \citet{Wilks2009} in
this setting. To put the method into context, forecasts for the
probability of the precipitation amount exceeding a certain threshold
have commonly been obtained using either quantile regression
\citep{Bremnes2004} or logistic regression (\citecs{WilksHamill2007};
Hamill, Hagedorn and Whitaker, \citeyear{Hamill2008}). If a full predictive distribution is sought,
such methods frequently fail, as they typically are inconsistent across
thresholds, violating the monotonicity constraint for cumulative
distribution functions. For quantile regression, \citet{Dette2012} and
\citet{Kneib2013} describe possible solutions to this problem. In the
case of the logistic regression approach, \citet{Wilks2009} proposes an
elegant remedy. In his method, the postprocessed predictive cumulative
distribution function takes the form
%
%
\begin{eqnarray}
\label{eqNRprecip}\qquad
&&G(y | x_1,\ldots, x_M) \nonumber\\[-8pt]\\[-8pt]
&&\quad=
\frac{\exp(a + b_1 x_1 + \cdots+ b_m x_M + h(y))} {
1 + \exp(a + b_1 x_1 + \cdots+ b_m x_M + h(y))},\nonumber
\end{eqnarray}
where $h$ grows strictly monotonically and without bounds as a
function of the precipitation accumulation $y \geq0$. Linear choices
for $h$ result in mixtures of a point mass at zero and a truncated
logistic distribution and, in light of the parametric family in
(\ref{eqNRprecip}), the technique can be interpreted as an NR
approach. More general formulations that allow for interaction terms
have recently been proposed by \citet{Ben2013}. As an alternative,
\citet{Scheuerer2013} introduces an NR approach in terms of
generalized extreme value (GEV) distributions.

\subsection{Wind} \label{secwind}

A wind vector can be represented by wind speed and wind direction or
by its $u$ (zonal or west--east) and $v$ (meridional or north--south)
velocity components. Wind speed is a nonnegative continuous variable.
\citet{SloughterGneitingRaftery2010} provide an ensemble BMA
implementation, where the kernel is a gamma distribution with the mean
and the variance being affine functions of the respective ensemble
member forecast. \citet{ThorarinsdottirGneiting2010} and
\citet{ThorarinsdottirJohnson2011} develop an NR approach in which the
predictive distribution is truncated normal. Wind direction is a
circular quantity and \citet{Bao2010} propose an ensemble BMA
specification where the kernel is a von Mises distribution.

When a wind vector is represented by its $u$ and $v$ components, the
methods described in Section~\ref{sectemperature} for temperature and
pressure become available, and examples of NR postprocessed predictive
distributions of the form (\ref{eqNRtemp}) for the $u$ component are
shown in Figure~\ref{figpredPDFwind}. In recent work, truly
bivariate postprocessing techniques for wind vectors have become
available, taking dependencies between the components into account
(\citecs{Pinson2011}; Schuhen, Thorarinsdottir and
Gneiting, \citeyear{Schuhen2012}; Sloughter, Gneiting and
Raftery, \citeyear{Sloughter2011}).
These methods are
discussed in subsequent sections.

\subsection{Estimation} \label{secestimation}

Ensemble postprocessing techniques depend on the availability of
training data for estimating the predictive model. Typically, optimum
score approaches have been used for estimation \citep{Gneiting2005},
with the maximum likelihood technique being a special case thereof
\citep{GneitingRaftery2007}, and Bayesian approaches offering
alternatives ({Di~Narzo and Cocchi}, \citeyear{DiNarzoCocchi2010}).

The training data are usually taken from a rolling training period
consisting of the recent past, including the most recent available
ensemble forecasts\break  along with the corresponding realizing values.
Common choices for the length of the training period range from 20 to
40 days. In schemes of this type, the training set is updated
continually, thereby allowing the estimates to adapt to changes in the
seasons and weather regimes. Clearly, there is a trade-off here, in
that larger training periods may allow for better estimation in
principle, thereby reducing estimation variances, but may introduce
biases due to seasonal effects. More flexible, adaptive estimation
approaches, such as recursive maximum likelihood techniques, have been
proposed and studied by \citet{Pinson2009}, \citet{Raftery2010} and
\citet{Pinson2011}.

In addition to deciding on the temporal extent of training sets,
choices regarding their spatial composition are to be made. Local
approaches use training data from the station location or grid box at
hand only, resulting in distinct sets of coefficients that are
tailored to the local terrain, while regional approaches composite
training sets spatially, to estimate a single set of coefficients that
is then used over an entire region
\citep{ThorarinsdottirGneiting2010}. Recently, flexible spatially
adaptive ap-\break proaches have been developed that estimate coefficients at
each station location individually, interpolating them to sites where
no observational assets are available (\citecs{Kleiber2011a};
Kleiber, Raftery and
Gneiting, \citeyear{Kleiber2011b}).

Introduced by \citeauthor{Hamill2006},\break (\citeyear{Hamill2006}), reforecasts are retrospective
weather forecasts with today's NWP models applied to past
initialization and valid dates. As reforecasts are based on the model
version that is currently run operationally, the availability of
reforecast data sets results in massive enlargements of training sets
for statistical postprocessing. The ensuing gains in the predictive
performance can be substantial, as demonstrated by
\citet{Hagedorn2008}, Hamill,  Hagedorn and
Whitaker, \citeyear{Hamill2008} and \citet{Hagedorn2012}, among
others.

\section{From Univariate to Multivariate Predictive Distributions:
Copula Approaches}
\label{seccopulas}

The univariate postprocessing methods discussed thus far yield
significant improvement in the predictive performance of raw NWP
ensemble output. However, in many applications it is critical that
multivariate dependencies in the forecast error, including the case of
temporal, spatial and spatio-temporal weather trajectories, are accounted
for. For example, winter road maintenance requires joint probabilistic
forecasts of temperature and precipitation \citep{Berrocal2010}, air
traffic control calls for probabilistic forecasts of wind fields
(Chaloulos and Lygeros, \citeyear{ChaloulosLygeros2007}), the management of renewable energy
resources hinges on spatio-temporal weather trajectories
\citep{Pinson2013}, and NWP output is used to drive hydrologic models
to address tasks such as flood warnings, the operation of waterways
and releases from reservoirs, with Schaake et al. [(\citeyear
{Schaake2010}), pages 61--62]
noting in this context that

\begin{quote}
``relationships between physically dependent variables like,
for example, precipitation and temperature should be respected.''
\end{quote}

\noindent
If statistical postprocessing proceeds independently for each weather
variable, location and look-ahead time, such relationships are ignored,
and it is critical that they be restored.

Toward this end, we recall Sklar's theorem, which is of fundamental
theoretical importance in dependence modeling, and we review Gaussian
and other parametric copulas approaches to the statistical
postprocessing of multivariate ensemble output. Then we turn to
empirical copulas, which permit the adoption of a rank order structure
from data records, as exemplified by the Schaake shuffle technique of
\citet{Clark2004}.

\subsection{Handling Dependencies: Sklar's Theorem} \label{secdependencies}

Taking a technical perspective momentarily, suppose that we have a
postprocessed predictive cumulative distribution function, $F_l$, for
each univariate weather quantity $Y_l$, where $l = 1,\ldots, L$, with
the multi-index $l = (i, j, k)$ referring to weather variable $i$,
location $j$ and look-ahead time $k$. What we seek is a physically
realistic multivariate joint predictive cumulative distribution
function $F$ with margins\break  $F_1,\ldots, F_L$.

Recall that a copula is a multivariate cumulative distribution
function with standard uniform margins (\citecs{Joe1997}; \citecs{Nelsen2006}).
Copulas have been employed successfully in a wealth of applications,
such as in finance (\citeauthor{McNeil2005},\break  \citeyear{McNeil2005}), hydrology
\citep{GenestFavre2007} and climatology
\citep{SchoelzelFriederichs2008}, to name but a few. Their relevance
stems from the following celebrated theorem of \citet{Sklar1959}.

%
%
\begin{theorem}[(Sklar)]
For any multivariate cumulative distribution function $F$ with
margins\break  $F_1,\ldots, F_L$ there exists a copula $C$ such that
%
%
\begin{equation}
\label{eqSklar} F(y_1,\ldots, y_L) = C
\bigl(F_1(y_1),\ldots, F_L(y_L)
\bigr)
\end{equation}
for $y_1,\ldots, y_L \in\real$. Furthermore, $C$ is unique on the
range of the margins.
\end{theorem}

In particular, Sklar's theorem demonstrates that univariate approaches
to the statistical postprocessing of ensemble output can accommodate
any type of joint dependence structure, provided that a suitable
copula function is specified. As copula methods allow for the
modeling of the marginal distributions and of the multivariate
dependence structure, as embodied by the copula, to be decoupled, they
are well suited for our problem.

\subsection{Gaussian and Other Parametric Copula Approaches}
\label{secparametric}

If the dimension $L$ of the output quantity is small, or if specific
structure can be exploited, such as in spatial or temporal settings,
parametric or semiparametric families of copulas can be employed.

The most common parametric approaches invoke a Gaussian copula
framework, under which the multivariate cumulative distribution
function $F$ is of the form
%
%
\begin{eqnarray}
\label{eqGaussian}\quad
&&C(y_1,\ldots, y_L | \Sigma) \nonumber\\[-8pt]\\[-8pt]
&&\quad=
\Phi_L \bigl( \Phi^{-1}\bigl(F_1(y_1)
\bigr),\ldots, \Phi^{-1}\bigl(F_L(y_L)\bigr) |
\Sigma\bigr),\nonumber
\end{eqnarray}
where $\Phi_L( \cdot| \Sigma)$ is the cumulative
distribution function of an $L$-variate normal distribution with mean
zero and correlation matrix $\Sigma$, and $\Phi^{-1}$ is the quantile
function of the univariate standard normal distribution. The use of
Gaussian copulas makes for a particularly tractable approach, as only
the correlation matrix $\Sigma$ needs to be modeled. In a recent
paper, \citet{Moeller2012} propose the use of Gaussian copulas to
recover the cross-variable dependence structure for multi-varia\-ble
forecasts at individual locations, where the ensemble BMA methodology
is used to obtain the postprocessed marginal predictive distributions.
The\break  method is straightforward except that precipitation requires
special treatment due to the mixed discrete-continuous nature of the
variable. The recent work of \citet{Pinson2011} and
\citet{Schuhen2012} on bivariate wind vectors invokes multivariate
normal predictive distributions, corresponding to the special case in
(\ref{eqGaussian}) in which the margins $F_1,\ldots, F_L$ are
normal.

The use of Gaussian copula methods has a long and well-established
tradition in geostatistics, where the approach is referred to as
anamorphosis; see \citet{ChilesDelfiner} and the references therein.
In the spatial setting, the correlation matrix $\Sigma$ in
(\ref{eqGaussian}) is taken to be highly structured, satisfying
assumptions such as spatial stationarity and/or isotro\-py, as
exemplified by \citeauthor{Gel2004}\break  (\citeyear{Gel2004}) and Berrocal, Raftery and
Gneiting (\citeyear{Berrocal2007,Berrocal2008}) in ensemble BMA approaches to temperature and
precipitation field forecasting. Similarly, Gaussian copulas have
been employed to capture dependencies over consecutive lead times in
postprocessed predictive distributions (\citecs{Pinson2009};
Schoel\-zel and Hense, \citeyear{SchoelzelHense2011}). When the margins $F_1,\ldots, F_L$ are normal,
the underlying stochastic model is that of a Gaussian process or
Gaussian random field, and choices in the parameterization of the
correlation matrix $\Sigma$ correspond to the selection of a
parametric correlation model in spatial statistics (\citecs{Stein};
\citecs{CressieWikle}).

%
%
\begin{figure*}

\includegraphics{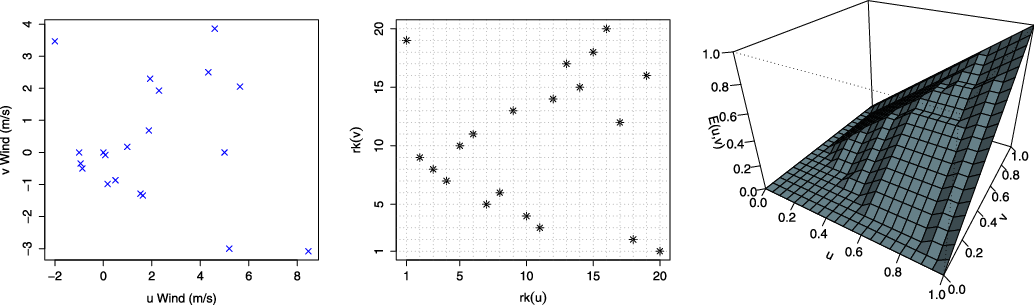}

\caption{The bivariate empirical distribution of the observed $u$ and
$v$ wind components at Hamburg at 2:00 am on April 1--20,
2011. Left: bivariate scatterplot. Middle: representation of the
rank dependence structure by a Latin square. Right: empirical
copula.} \label{figSchaake}
\end{figure*}
%

While Gaussian copulas yield convenient, ubiquitous stochastic models,
parametric or semiparametric alternatives are available, including but
not limited to the use of elliptical copulas
({Demarta and McNeil}, \citeyear{DemartaMcNeil2005}), Archimedian copulas
({McNeil and
Ne{\v{s}}lehov{\'a}}, \citeyear{McNeilNeslehova2009}),
extremal copulas ({Davison, Padoan and
Ribatet}, \citeyear{Davison2012})
and pair copulas ({Aas et~al.}, \citeyear{Aas2009}).

\subsection{Empirical Copulas} \label{secempirical}

In the common case in which the dimension $L$ of the output quantity
is huge and no specific structure can be exploited, parametric methods
are bound to fail. We then need to resort to nonparametric approaches
that depend on the use of empirical copulas. Here, let $\{
(x_m^1,\ldots, x_m^L)\dvtx m = 1,\ldots,M \}$ denote a data set of size
$M$ with values in $\real^L$. Assuming for simplicity that there are
no ties, let $\operatorname{rk}(x_m^l)$ denote the rank of $x_m^l$
within $x_1^l,\ldots, x_M^l$. The corresponding empirical copula
$E_M$ is defined as
%
%
\begin{eqnarray}
\label{eqempirical}\quad
&&E_M \biggl( \frac{i_1}{M},\ldots,
\frac{i_L}{M} \biggr) \nonumber\\[-8pt]\\[-8pt]
&&\quad= \frac{1}{M} \sum_{m=1}^M
\mathbh{1} \bigl\{ \operatorname{rk}\bigl(x_m^1\bigr)
\leq i_{1},\ldots, \operatorname{rk}\bigl(x_m^L
\bigr) \leq i_{L} \bigr\}\nonumber
\end{eqnarray}
for integers $0 \leq i_1,\ldots, i_L \leq M$; see
\citet{Deheuvels1979}, who uses the term empirical dependence
function, and \citet{Rueschendorf2009} and the references therein.

Any empirical copula is an irreducible discrete copula in the sense
described by \citet{Koles2006}, with \citet{Mayor2007} providing a
bivariate version of Sklar's theorem in this setting. As we will
illustrate below, empirical copulas can be thought of as corresponding
to Latin hypersquares. Asymptotic theory for the respective empirical
processes has been developed by R{\"u}schen\-dorf
(\citeyear{Rueschendorf1976,Rueschendorf2009}), \citet{Stute1984}, \citet{VanderVaartWellner1996},
\citet{Fermanian2004} and \citet{Segers2012}, among other authors.

In the context of nonparametric approaches to the statistical
postprocessing of multivariate NWP ensemble output, empirical copulas
allow for the adoption of a multivariate rank order structure either
from historical weather observations, as in the Schaake shuffle
technique of \citet{Clark2004}, or directly from the ensemble
forecast, to be discussed in detail in Section~\ref{secECC}.

\subsection{The Schaake Shuffle} \label{secSchaake}

Clark et~al. (\citeyear{Clark2004}) introduced the ingenious\break  Schaake shuffle as a method
for reconstructing physically realistic spatio-temporal structure in
forecast\-ed temperature and precipitation fields. Even though it has
been presented as a reordering technique in the extant literature, an
empirical copula interpretation of the Schaake shuffle is readily
available.

Consider an output quantity taking values in $\real^L$ and suppose
that we have univariate postprocessed predictive distributions
$F_1,\ldots, F_L$ for the margins. Suppose, furthermore, that we have a set
of $M$ historical weather field observations for the $\real^L$-valued
output quantity at hand. From the historical record, we can construct
an empirical copula of the form (\ref{eqempirical}), as illustrated
in the right-hand panel of Figure~\ref{figSchaake}, where we merely
have $L = 2$ as corresponds to the components of a wind vector and $M
= 20$.

%
%
\begin{figure*}

\includegraphics{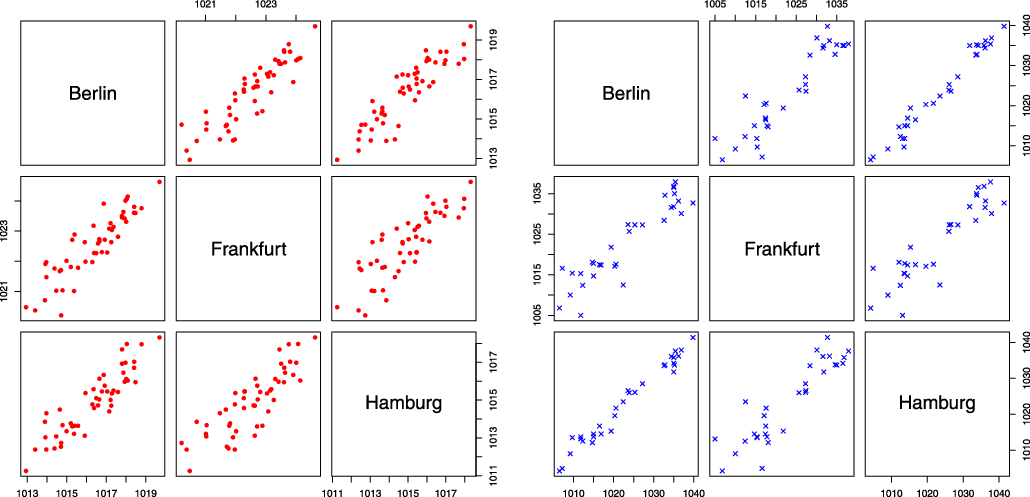}

\caption{Scatterplot matrices for pressure at Berlin, Frankfurt and
Hamburg. Left: 48-hour ahead ECMWF ensemble forecast valid 2:00 am
on April 1, 2011. Right: empirical distribution of the pressure
observations at the same hour over the period March 1--31, 2011.}
\label{figmotivation}
\end{figure*}
%

To apply the Schaake shuffle, we take a discrete sample of size $M$
from each of the univariate postprocessed predictive distributions
$F_1,\ldots, F_L$, and\break  then we reorder to match with the rank order
structure in the historical record, which is also of size $M$. This
procedure corresponds to the application of the empirical copula of
the historical weather field record to the discrete samples from the
univariate postprocessed predictive distribution, and in this sense it
is natural to consider the Schaake shuffle as an empirical copula
technique. The thus reordered forecast inherits the multivariate rank
dependence structure and the pairwise Spearman rank correlation
coefficients from the historical weather record at hand. A more
technical discussion can be given in close analogy to what we describe
in Section~\ref{secinterpretation} within the related context of the
ensemble copula coupling approach.

The Schaake shuffle has met great success in meteorological and
hydrologic applications, where it recovers observed spatial and
cross-variable dependence structures as well as temporal
persistence\break
(\citecs{Clark2004}; \citecs{Schaake2007}; \citecs{Voisin2011}).
Nevertheless, there is
a major limitation, in that the standard implementation fails to
condition the multivariate dependence structure on current or
predicted atmospheric conditions. Clark et al. [(\citeyear{Clark2004}),
page 260]
therefore describe a future extension of the Schaake shuffle, the idea
of which is as follows:

\begin{quote}
``to preferentially select dates from the historical record that resemble
forecasted atmospheric conditions and use the spatial correlation structure
from this subset of dates to reconstruct the spatial variability for a
specific forecast.''
\end{quote}

In what follows we pursue a related empirical copula approach, in
which the postprocessed forecast inherits the multivariate dependence
structure from the raw NWP ensemble, rather than from a historical
record of weather observations, thereby addressing the lack of
atmospheric flow and time dependence in the standard Schaake shuffle.

\section{Ensemble Copula Coupling (ECC)} \label{secECC}

The ensemble copula coupling (ECC) approach draws on the rank order
information available in the raw ensemble forecast, based on the
implicit assumptions that its members are exchangeable and that the
NWP ensemble is capable of representing observed cross-variable,
spatial and temporal dependence structures. While the latter is to be
expected, given that NWP models discretize the equations that govern
the physics of the atmosphere, diagnostic checks are advisable, to
assess empirically whether dependence structures in individual
ensemble forecasts are compatible with observational re-\break cords. We give
a simple illustration in Figure~\ref{figmotivation}, where the
dependence structures within the ensemble forecast valid April 1, 2011
and those in the observational record over the preceding month
resemble each other strongly.

\subsection{The ECC Approach} \label{secapproach}

The ECC approach is a general multi-stage procedure for the generation
of a postprocessed ensemble of the same size, $M$, as the raw
ensemble. We write $x_1^l,\ldots, x_M^l$ for the univariate margins
of the raw ensemble, where the multi-index $l = (i,j,k)$ refers to
weather variable $i \in\{ 1,\ldots, I \}$, location $j \in\{ 1,\ldots,
J \}$ and lead time $k \in\{ 1,\ldots, K \}$, to comprise
NWP output in $\real^L$, where the dimension is $L = I \times J \times
K$. In order to generate an ECC postprocessed ensemble forecast, we
proceed as follows.

\begin{longlist}
\item[\textit{Univariate postprocessing}.]
For each margin $l$, obtain a postprocessed predictive distribution,
$F_l$, by applying a univariate postprocessing technique, such as
ensemble BMA or NR, to the raw ensemble output
%
%
\begin{equation}
\label{eqrawens} x_1^l,\ldots, x_M^l.
\end{equation}

\item[\textit{Quantization}.]
Represent each univariate predictive distribution $F_l$ by a discrete
sample of size $M$, say,
%
%
\begin{equation}
\label{eqindependentens} \tilde{x}_1^l,\ldots,
\tilde{x}_M^l.
\end{equation}
The discrete sample can be generated in various ways, to be discussed
in detail in Section~\ref{secquantization}, where we distinguish the
ECC-Q, ECC-R and ECC-T va\-riants, depending on how the quantization is
performed.\footnote{Note that the quantized values in
(\ref{eqindependentens}) may be ordered, as in the case of the ECC-Q
approach, or may not be ordered, as in the case of the ECC-R and
ECC-T scheme, respectively.}

\item[\textit{Ensemble reordering}.]
For each margin $l$, the order statistics\footnote{The $k$th order
statistic of a sample is defined as its $k$th smallest value. For
each margin $l$, we write $x_{(1)}^l \leq\cdots\leq x_{(M)}^l$ and
$\tilde{x}_{(1)}^l \leq\cdots\leq\tilde{x}_{(M)}^l$ for the order
statistics\vspace*{1pt} of the raw ensemble values in (\ref{eqrawens}) and the
quantized values in (\ref{eqindependentens}), respectively. The
latter appear on the right-hand side of (\ref{eqECCens}), where
we define the ECC postprocessed ensemble.} of the
raw ensemble values,
\[
x^l_{(1)} \leq\cdots\leq x^l_{(M)}
\]
induce a permutation $\sigma_l$ of the integers $\{ 1,\ldots, M
\}$, defined by $\sigma_l(m) = \operatorname{rk} (x_m^l)$ for $m =
1,\ldots, M$. If there are ties among the ensemble values, the
corresponding ranks can be allocated at random.\footnote{While
randomization is a natural approach in the case of ties, other allocation
methods are feasible and do not pose technical problems. Regardless of the
allocation, equation (\ref{eqempirical}) continues to apply.} The respective
margin of the ECC postprocessed ensemble is then given by
%
%
\begin{equation}
\label{eqECCens} \hat{x}^l_1 = \tilde{x}^l_{(\sigma_l(1))},\ldots, \hat
{x}^l_M = \tilde{x}^l_{(\sigma_l(M))}.
\end{equation}
\end{longlist}

Note that, while the permutation $\sigma_l$ is determined by the order
statistics of the raw ensemble, equation (\ref{eqECCens}) applies this
permutation to the postprocessed and quantized values.

The ECC approach is attractive computationally, in that the modeling
of the multivariate dependence structure requires only the calculation
of marginal ranks. In the recent literature, the approach has been
introduced as a reordering technique, as described colorfully by
\citet{Flowerdew2012}, page 15:

\begin{quote}
``The key to preserving spatial, temporal and inter-variable structure
is how this set of values is distributed between ensemble members. One
can always construct ensemble members by sampling from the calibrated
PDF, but this alone would produce spatially noisy fields lacking the
correct correlations. Instead, the values are assigned to ensemble
members in the same order as the values from the raw ensemble: the
member with the locally highest rainfall remains locally highest, but
with a calibrated rainfall magnitude.''
\end{quote}

\noindent
That said, it is fruitful to interpret the ECC approach as a
nonparametric copula technique, which permits us to fuse and
consolidate seemingly unrelated, recent advances within a single,
structured framework.

\subsection{Empirical Copula Interpretation} \label{secinterpretation}

Elaborating on our interpretation of the Schaake shuffle, we now
demonstrate that the ECC approach can be considered an empirical
copula technique. For convenience, we assume that there are no ties
among the raw ensemble margins. We write $R_1,\ldots, R_L$ for the
corresponding marginal empirical cumulative distribution functions,
which take values in the set
\[
I_M = \biggl\{ 0, \frac{1}{M},\ldots, \frac{M-1}{M}, 1
\biggr\}.
\]
The multivariate empirical cumulative distribution function $R\dvtx
\real^L \to I_M$ of the raw ensemble maps\break  into~$I_M$, too. According
to the discrete version of Sklar's theorem described by
\citet{Mayor2007} in the bivariate case, there exists a unique\-ly
determined empirical copula $E_M\dvtx I_M^L \to I_M$ such that
%
%
\begin{equation}
\label{eqrawmCDF}\quad R(y_1,\ldots, y_L) = E_M
\bigl( R_1(y_1),\ldots, R_L(y_L)
\bigr)
\end{equation}
for all $y_1,\ldots, y_L \in\real$, allowing for the same type of
interpretation as illustrated in Figure~\ref{figSchaake} in the case
of the Schaake shuffle.

Analogous considerations apply to the quantized independently
postprocessed ensemble (\ref{eqindependentens}) and the ECC
postprocessed ensemble (\ref{eqECCens}). Using obvious notation, we
write $\tilde{F}$ and $\hat{F}$ for the corresponding multivariate
empirical cumulative distribution functions. Furthermore, we denote
the marginal empirical cumulative distribution functions of the
quantized independently postprocessed ensemble by\break  $\tilde{F}_1,\ldots,
\tilde{F}_L$, respectively, and we use the symbol
$\tilde{E}_M$ to denote the corresponding copula. Then
%
%
\begin{equation}
\label{eqindependentmCDF}\quad \tilde{F}(y_1,\ldots, y_L) =
\tilde{E}_M \bigl(\tilde{F}_1(y_1),\ldots,
\tilde{F}_L(y_L)\bigr)
\end{equation}
and
%
%
\begin{equation}
\label{eqECCmCDF}\quad \hat{F}(y_1,\ldots, y_L) =
E_M \bigl(\tilde{F}_1(y_1),\ldots,
\tilde{F}_L(y_L)\bigr)
\end{equation}
for all $y_1,\ldots, y_L \in\real$. As elucidated by equations
(\ref{eqrawmCDF}), (\ref{eqindependentmCDF}) and
(\ref{eqECCmCDF}), the quantized independently postprocessed
ensemble and the ECC postprocessed ensemble share the margins, whereas
the raw ensemble and the ECC postprocessed ensemble share the copula,
as illustrated in Figure~\ref{figECC}. In particular, the ECC
postprocessed ensemble honors and retains the flow-dependent
multivariate rank dependence structure and bivariate Spearman rank
correlation coefficients in the raw NWP ensemble output.

\subsection{ECC-Q, ECC-R and ECC-T} \label{secquantization}

We now discuss options for the generation of the discrete samples
(\ref{eqindependentens}) at the quantization stage of the ECC
approach. Perhaps the most natural way of obtaining a discrete sample
of size $M$ from the postprocessed predictive cumulative distribution
function $F_l$ is to take equidistant Quantiles of the form
%
%
\renewcommand{\theequation}{\mbox{ECC-Q}}
\begin{eqnarray}\label{eqECC-Q} \tilde{x}_1^l &=& F_l^{-1} \biggl( \frac{1}{M+1}
\biggr),\ldots,
\tilde{x}_M^l = F_l^{-1} \biggl( \frac{M}{M+1} \biggr),\nonumber
\\
\end{eqnarray}
and we refer to this approach as ECC-Q.\footnote{\citet{Broecker2012}
provides theoretical arguments in support of the particular choice of
the quantiles in (\ref{eqECC-Q}), which maintains the calibration of the
univariate ensemble forecasts, well in line with the goal of
maximizing the sharpness of the predictive distributions subject to
calibration \citep{GneitingBalabdaouiRaftery2007}. An alternative
choice would be to set
{\fontsize{8pt}{\baselineskip}\selectfont{
\begin{eqnarray*}
\tilde{x}_1^l &=& F_l^{-1} \biggl(
\frac{{1}/{2}}{M} \biggr),
\tilde{x}_2^l =
F_l^{-1} \biggl( \frac{{3}/{2}}{M} \biggr),\ldots,
\tilde{x}_M^l = F_l^{-1} \biggl(
\frac{M\!-\!{1}/{2}}{M} \biggr),
\end{eqnarray*}}}
which fails to maintain calibration in some respects, but is optimal
in expectation if the predictive performance is measured by the
continuous ranked probability score \citep{Broecker2012}. Related
optimality results can be found in the literature on the quantization
of probability distributions as reviewed by \citet{GrafLuschgy2000}.}
Another option is to take a simple Random sample of the form
%
%
\renewcommand{\theequation}{\mbox{ECC-R}}
\begin{equation}
\label{eqECC-R}  \tilde{x}_1^l =
F_l^{-1}(u_1),\ldots, \tilde{x}_M^l
= F_l^{-1}(u_M),\hspace*{-30pt}
\end{equation}
where $u_1,\ldots, u_M$ are independent standard uniform random variates.
We refer to this latter option as \mbox{ECC-R}.

Finally, we consider a quantile mapping or transformation approach
that generalizes a recent proposal by \citet{Pinson2011} in the case
of wind vectors. In this technique, we adopt the ensemble smoothing
approach of \citet{Wilks2002} and fit a parametric, continuous cumulative
distribution function $S_l$ to the raw ensemble margin $R_l$. We then
extract the quantiles from $F_l$ that correspond to the percentiles of
the raw ensemble values in $S_l$, in that
%
%
\renewcommand{\theequation}{\mbox{ECC-T}}
\begin{eqnarray}
\label{eqECC-T} \tilde{x}_1^l &=&
F_l^{-1}\bigl(S_l\bigl(x_1^l \bigr)\bigr),\ldots,
\tilde{x}_M^l = F_l^{-1}
\bigl(S_l\bigl(x_M^l\bigr)\bigr).\nonumber
\\
\end{eqnarray}
We refer to this Transformation approach for continuous variables as
ECC-T. Frequently, as in the case of temperature, pressure and the
$u$ and $v$ wind vector components, $S_l$ can be taken to be normal,
with mean equal to the ensemble mean and variance equal to the
ensemble variance. In the special situation in which $S_l$ and $F_l$
belong to the same location-scale family, such that $S_l(x) =
G((x-\mu)/\sigma)$ and $F_l(x) = G((x-\tilde\mu)/\tilde\sigma)$ for
some continuous cumulative distribution function~$G$, $\mu, \tilde\mu
\in\real$ and $\sigma, \tilde\sigma> 0$, the transformation from $x$
to
%
%
\setcounter{equation}{6}
\renewcommand{\theequation}{\arabic{section}.\arabic{equation}}
\begin{equation}
\label{eqECC-Ttransformation} \tilde{x} = F_l^{-1}
\bigl(S_l(x)\bigr) = \tilde\mu+ \frac{\tilde\sigma}{\sigma} ( x - \mu)
\end{equation}
becomes affine and, thus, the ECC-T postprocessed ensemble conserves
the raw ensemble's bivariate Pearson product moment correlation
coefficients, in addition to retaining its bivariate Spearman rank
correlation coefficients.

The discussion in \citet{Broecker2012} provides theoretical support in
favor of the ECC-Q approach, and so does our case study in Section~\ref{secresultsunivariate}, where we compare the predictive
performance of the ECC-Q, ECC-R and ECC-T schemes. We therefore
recommend the use of the natural ECC-Q approach.

\subsection{Relationships to Extant Work} \label{secrelations}

While the broad framework and the interpretation in terms of empirical
copulas in our paper are original, the idea of the ECC approach is not
new, with its recent appearances in the literature coming in various
seemingly unrelated shades and flavors. In this context, the
connections to the work of \citet{Pinson2011} and
\citet{RoulinVannitsem2012} are of particular interest.

The method described in Section~2.c of \citet{RoulinVannitsem2012} in
the context of areal precipitation forecasts can be viewed as a
variant of the ECC-Q scheme, as it extracts equally spaced quantiles
from the postprocessed marginal predictive cumulative distribution
functions, which are of logistic type, followed by a reordering with
respect to the raw ensemble values, with adaptations to account for a
point mass at zero.

\citet{Pinson2011} proposes a transformation technique for the
postprocessing of ensemble forecasts of wind vector components. In
this method, each postprocessed margin is a translated and dilated
version of the original margin, with the mapping being compatible with
the ECC-T scheme in the special case in which both $S_l$ and $F_l$ are
normal.

\section{Case Study} \label{secdata}

In this case study we exemplify the use of statistical postprocessing
techniques, illustrate and assess the ECC approach, and compare the
predictive performance of the ECC-Q, ECC-R and ECC-T schemes,
respectively. All forecasts are based on the 50-member global NWP
ensemble managed by the European Centre for Medium-Range Weather
Forecasts (ECMWF), which operates at a horizontal resolution of
approximately 32 km and lead times up to ten days ahead
(\citecs{Molteni1996}; {Leutbecher and
Palmer, \citeyear{LeutbecherPalmer2008}). The differences between the ensemble members
stem from random perturbations in initial conditions and stochastic
physics parameterizations and, thus, the ensemble members are
statistically indistinguishable and can be considered as exchangeable.

\subsection{Setting} \label{secsetting}

We restrict attention to the ECMWF ensemble run initialized at 00:00
Universal Time Coordinated (UTC) and consider forecasts for surface
temperature, sea level pressure, precipitation and the $u$ wind vector
component at lead times of 24 and 48 hours, with emphasis on the
international airports at Berlin--Tegel, Frankfurt am Main and Hamburg
in Germany, where 00:00 UTC corresponds to 2:00 am local time in
summer and 1:00 am local time in winter. The locations of the three
airports are marked in the upper left panel in Figure~\ref{figenstemp}. Our test period consists of the twelve month
period ranging from May 1, 2010 through April 30, 2011. Forecasts and
observations prior to May 1, 2010 are used as training data as needed.

To obtain postprocessed marginal predictive distributions for each
weather variable, location and lead time individually, we apply the
techniques described in Section~\ref{secunivariate}. For temperature
and pressure, we employ the ensemble BMA model (\ref{eqBMAtemp})
with a normal kernel, and for precipitation the Bernoulli--Gamma
ensemble BMA model specified in (\ref{eqBMAprecip0}),
(\ref{eqBMAprecip1}) and (\ref{eqBMAprecip2}), respectively.
For the wind vector components, we use the NR model
(\ref{eqNRtemp}). To fit the univariate predictive models, we use
local data from a rolling training period consisting of the most
recent available 30 days and employ the estimation techniques proposed
by \citet{Raftery2005}, \citet{Sloughter2007} and
\citet{Gneiting2005}. Then we apply the ECC-Q, ECC-R and ECC-T
schemes as described in Section~\ref{secECC}.

\subsection{Evaluation Methods} \label{secevaluation}

Statistical postprocessing techniques aim at generating calibrated and
sharp probabilistic forecasts from NWP ensemble output. As argued by
\citet{GneitingBalabdaouiRaftery2007}, the goal in probabilistic
forecasting is to maximize the sharpness of the predictive
distributions subject to calibration. Calibration is a multi-faceted,
joint property of the forecasts and the observations; essentially, the
forecasts are calibrated if the observations can be interpreted as
random draws from the predictive distributions. Sharpness refers to
the concentration of the predictive distributions, and thus is a
property of the forecasts only.

In univariate settings, calibration is checked via the probability
integral transform (PIT) or the verification rank. The PIT is simply
the value that the predictive cumulative distribution function attains
at the realizing observation (\citecs{Dawid1984}; Gneiting, Balabdaoui and
Raftery, \citeyear{GneitingBalabdaouiRaftery2007}), with suitable adaptations in
the case
of discrete distributions ({Czado, Gneiting and
Held}, \citeyear{CzadoGneitingHeld2009}). For an
ensemble forecast, the verification rank is the rank of the realizing
observation when pooled with the ensemble values \citep{Hamill2001}.
When a predictive distribution is calibrated, the PIT or verification
rank is uniformly distributed. Thus, calibration can be diagnosed by
compositing over forecast cases, plotting a PIT or verification rank
histogram, respectively, and checking for deviations from uniformity.
Verification rank and PIT histograms are directly comparable, with a
\mbox{U-shape} indicating underdispersion, an inverse \mbox{U-shape} indicating
overdispersion, and skew pointing at biases in the predictive
distributions.

Proper scoring rules provide decision theoretically coherent numerical
measures of predictive performance that may assess calibration and
sharpness simultaneously. Here we use the proper continuous ranked
probability score (CRPS), defined by
%
%
\begin{eqnarray}
\label{eqcrps}\quad \operatorname{crps}(F,y) & = & \int_{-\infty}^\infty
\bigl( F(z) - \mathbh{1} \{ y \leq z \} \bigr)^2 \,\mathrm{d} z
\\
\label{eqcrpskernel} & = & \mathbb{E}_F |X - y| -
\frac{1}{2} \mathbb{E}_F \bigl|X - X'\bigr|,
\end{eqnarray}
where $F$ is a predictive cumulative distribution function with finite
first moment, $y$ is the verifying observation, and $X$ and $X'$ are
independent random variables with distribution $F$
(\citeauthor{GneitingRaftery2007},\break  \citeyear{GneitingRaftery2007}). If $F$ corresponds to a point measure
$\delta_x$, the proper continuous ranked probability score reduces to
the absolute error, $|x - y|$. If $F = F_{\mathrm{ens}}$ is an ensemble
forecast with members $x_1,\ldots, x_M \in\real$, we interpret it as
an empirical measure and compute the continuous ranked probability
score as
%
%
\begin{eqnarray}
\label{eqcrpsens} \operatorname{crps}(F_{\mathrm{ens}},y) &=& \frac{1}{M}
\sum_{m=1}^M |x_{m} - y| \nonumber\\[-8pt]\\[-8pt]
&&{}-
\frac{1}{2M^{2}} \sum_{n=1}^M \sum
_{m=1}^M |x_n -
x_m|.\nonumber
\end{eqnarray}
We furthermore find the absolute error for the point forecast given by
the median of the predictive distribution, which is the Bayes
predictor under this loss function \citep{Gneiting2011}. Forecasting
methods then are compared by averaging scores over the test set, with
smaller values indicating better predictive performance.

To assess the calibration of ensemble forecasts of a multivariate
quantity, we use the multivariate version of the rank histogram
described by \citet{Gneiting2008}. We also employ the proper energy
score, which generalizes the continuous ranked probability score in the
representation (\ref{eqcrpskernel}), and is defined as
%
%
\begin{equation}
\label{eqes}\quad \operatorname{es}(F,y) = \mathbb{E}_F \| X - y \| -
\tfrac{1}{2} \mathbb{E}_F \bigl\| X - X' \bigr\|,
\end{equation}
where $\| \cdot\|$ denotes the Euclidean norm, $F$ is a predictive
distribution with finite first moments, $X$ and $X'$ are independent
random vectors with distribution $F$, and $y$ is the verifying
observation ({Gneiting and
Raftery}, \citeyear{GneitingRaftery2007}). For ensemble forecasts the
natural analogue of the formula (\ref{eqcrpsens}) applies. If the
scales of the weather variables vary, the margins should be
standardized before computing the joint energy score for these
variables. This can be done using the marginal means and standard
deviations of the observations in the test set.

The aforementioned techniques for the evaluation of probabilistic
forecasts of multivariate quantities have been developed with
low-dimensional quantities in mind \citep{Gneiting2008}, and we apply
them in dimension $L \leq3$ only. In higher dimension, these methods
lose power, and there is a pronounced need for the development of
theoretically principled evaluation techniques that are tailored to
such settings (\citecs{Pinson2013}, Section~5.2).

%
%
\begin{table*}[t]
\caption{Mean continuous ranked probability score (CRPS) and mean
absolute error (MAE) for univariate forecasts of temperature,
pressure, precipitation and the $u$ wind component at Berlin,
Frankfurt and Hamburg, at lead times of 24 and 48 hours,
respectively, for a test period ranging from May 1, 2010 through
April 30, 2011}
\label{tabunivariate}
\begin{tabular*}{\tablewidth}{@{\extracolsep{4in minus 4in}}lcccccccc@{}}
\hline
&&& \multicolumn{3}{c}{\textbf{CRPS}} & \multicolumn{3}{c@{}}{\textbf{MAE}} \\[-4pt]
&&& \multicolumn{3}{l}{\rule{158pt}{1pt}} & \multicolumn{3}{r@{}}{\rule{158pt}{1pt}}
\\
&&& \textbf{Berlin} & \textbf{Frankfurt} & \textbf{Hamburg}
& \textbf{Berlin} & \textbf{Frankfurt} & \textbf{Hamburg} \\
\hline
Temp. & 24 & ECMWF & $1.21$ & $1.23$ & $1.01$ & $1.50$ & $1.53$ &
$1.26$ \\
($^\circ$C) & & BMA & $0.90$ & $0.88$ & $0.79$ & $1.27$ & $1.23$ &
$1.10$ \\
[2pt]
& 48 & ECMWF & $1.25$ & $1.26$ & $1.06$ & $1.62$ & $1.62$ & $1.39$ \\
& & BMA & $0.99$ & $0.97$ & $0.92$ & $1.41$ & $1.33$ & $1.31$ \\
[6pt]
Pressure & 24 & ECMWF & $0.54$ & $0.55$ & $0.51$ & $0.75$
& $0.75$ & $0.71$ \\
(hPa) & & BMA & $0.43$ & $0.43$ & $0.39$ & $0.62$ & $0.61$ & $0.54$ \\
[2pt]
& 48 & ECMWF & $0.80$ & $0.78$ & $0.77$ & $1.12$ & $1.08$ & $1.09$ \\
& & BMA & $0.77$ & $0.74$ & $0.73$ & $1.08$ & $1.03$ & $1.03$ \\
[6pt]
Precip. & 24 & ECMWF & $0.25$ & $0.41$ & $0.31$ & $0.32$ & $0.51$ &
$0.39$ \\
(mm) & & BMA & $0.23$ & $0.40$ & $0.37$ & $0.30$ & $0.49$ & $0.44$ \\
[2pt]
& 48 & ECMWF & $0.26$ & $0.41$ & $0.36$ & $0.34$ & $0.50$ & $0.45$ \\
& & BMA & $0.26$ & $0.43$ & $0.39$ & $0.32$ & $0.52$ & $0.48$ \\
[6pt]
$u$ Wind & 24 & ECMWF & $0.83$ & $0.96$ & $0.89$ & $1.06$
& $1.19$ & $1.11$ \\
(m/s) & & NR & $0.70$ & $0.60$ & $0.68$ & $0.97$ & $0.81$ & $0.96$ \\
[2pt]
& 48 & ECMWF & $0.82$ & $0.89$ & $0.88$ & $1.09$ & $1.15$ & $1.18$ \\
& & NR & $0.75$ & $0.62$ & $0.75$ & $1.05$ & $0.83$ & $1.04$ \\
\hline
\end{tabular*}
\end{table*}

%
%
\begin{figure*}[t]

\includegraphics{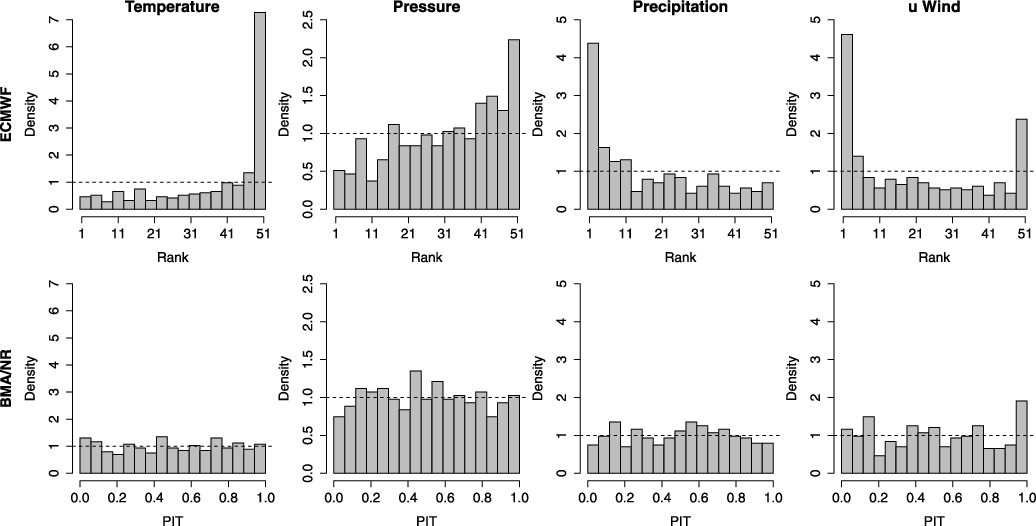}

\caption{Calibration checks for 48-hour ahead forecasts of
temperature, pressure, precipitation and $u$ wind at Frankfurt, for
a test period ranging from May 1, 2010 through April 30, 2011. Top:
verification rank histograms for the ECMWF ensemble. Bottom: PIT
histograms for BMA or NR postprocessed predictive
distributions.}
\label{figRH}
\end{figure*}
%

\subsection{Predictive Performance for Univariate Weather Quantities}
\label{secresultsunivariate}

Table~\ref{tabunivariate} compares the predictive performance of the
raw ECMWF ensemble and the postprocessed predictive distributions for
temperature, pressure, precipitation and the $u$ wind vector component
at lead times of 24 and 48 hours at Berlin, Frankfurt and Hamburg,
respectively. The BMA and NR postprocessing generally leads to a
significant improvement in the predictive skill, as measured by the
mean CRPS and the MAE, with exceptions in the case of
precipitation.\footnote{The particularly good performance of the raw
ensemble for precipitation accumulations at the stations considered
and potential shortcomings in the details of the postprocessing
technique \citep{Scheuerer2013} may serve to explain these
exceptions.} Not unexpectedly, the performance generally is better at
the shorter prediction horizon of 24 hours.

Figure~\ref{figRH} shows verification rank and PIT histo\-grams for
temperature, pressure, precipitation and $u$ wind at a lead time of 48
hours at Frankfurt. The postprocessed forecasts show much better
calibration, as evidenced by the nearly uniform PIT histograms, except
perhaps in the case of precipitation, where a slight inverse U-shape
of the PIT histogram may indicate overdispersion in the BMA
postprocessed predictive distributions.

\subsection{Predictive Performance for Multivariate Weather Quantities}
\label{secresultsmultivariate}

We now give an illustration and initial evaluation of ECC
postprocessed multivariate predictive distributions.

Table~\ref{tabspatial} and Figure~\ref{figspatial} concern
temperature and pressure, with each of these variables being
considered at Berlin, Frankfurt and Hamburg jointly. The distance
from Frankfurt to either Berlin or Hamburg is on the order of 400
kilometers, and the distance between Berlin and Hamburg is
approximately 250 kilometers. Wind and precipitation patterns vary at
considerably smaller spatial scales and we thus do not expect ECC to
make much of a difference here. In contrast, forecast errors\ for
pressure can be expected to show pronounced long range dependencies,
and perhaps to some lesser extent for temperature. The scores and
multivariate rank histograms confirm the strongly positive effects of
ECC in the case of pressure, where the ECC postprocessed tri\-variate
predictive distributions are much better calibrated than either the
raw ensemble or the independent BMA postprocessed predictive
distributions. The ECC-Q quantization scheme outperforms the ECC-R
and ECC-T approaches.

%
%
\begin{table}
\caption{Mean energy score for 48-h ahead forecasts of temperature and
pressure, each considered at Berlin, Frankfurt and Hamburg jointly,
for a test period ranging from May 1, 2010 through April 30, 2011.
The scores for the independent BMA and ECC-R techniques, which
involve randomization, are averaged over 100~repetitions}
\label{tabspatial}
\begin{tabular*}{\tablewidth}{@{\extracolsep{\fill}}lcc@{}}
\hline
& \textbf{Temperature} & \textbf{Pressure} \\
& \textbf{($\bolds{^\circ}$C)} & \textbf{(hPa)} \\
\hline
ECMWF & 2.342 & 1.478 \\
BMA & 1.929 & 1.473 \\
ECC-Q & 1.927 & 1.428 \\
ECC-R & 1.945 & 1.454 \\
ECC-T & 1.934 & 1.442 \\
\hline
\end{tabular*}
\end{table}
%

%
%
\begin{figure*}[t]

\includegraphics{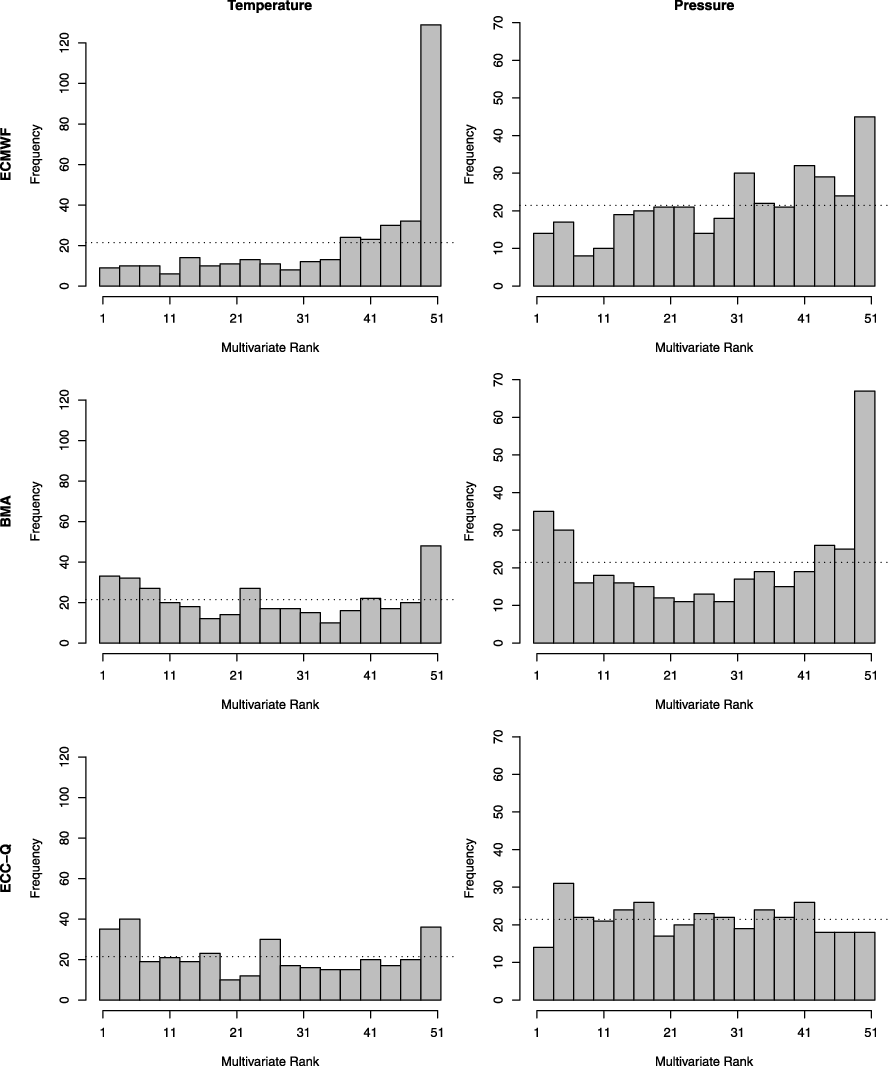}

\caption{Multivariate rank histograms for 48-h ahead ensemble
forecasts of temperature and pressure, each considered at Berlin,
Frankfurt and Hamburg jointly, for a test period ranging from May 1,
2010 through April 30, 2011.}\label{figspatial}
\end{figure*}
%

While for temperature the BMA postprocessing improves strongly on the
raw ensemble forecast, the effect of ECC is minor, if not negative,
due to the correlations in the forecast errors being negligible at the
distances considered here. That said, Figure~\ref{figspatialtemp}
illustrates the strongly positive effects of ECC on temperature field
forecasts, where dependencies at short and moderate distances are of
critical importance. Here we consider $33 \times37 = 1221$ NWP
model grid boxes over Germany and adjacent areas, with the forecast
made a day ahead for 2:00 am on April 25, 2011, for what promises to
be a pleasant, unusually warm spring night.

%
%
\begin{figure*}

\includegraphics{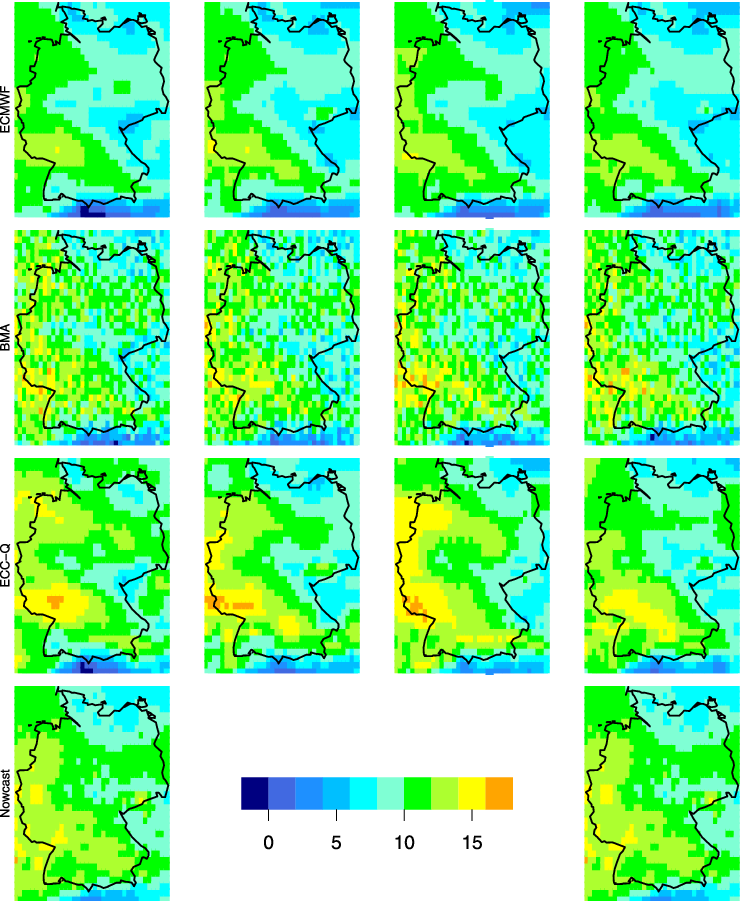}
 \caption{24-hour ahead ensemble forecasts for
temperature over Germany valid 2:00 am on April 25, 2011, in the unit
of degrees Celsius. Top row: four randomly selected members of the raw
ECMWF ensemble. Second row: independent BMA postprocessing---for each
grid box, a~random number from the corresponding BMA postprocessed
predictive distribution is drawn. Third row: four members of the
corresponding ECC ensemble, with rank order structures adopted from the
respective raw ensemble members in the top row. Bottom row:
single-valued nowcast as described in the text, shown both at left and
at right.} \label{figspatialtemp}
\end{figure*}
%

The postprocessing uses a single BMA model of the form
(\ref{eqBMAtemp}), which is trained on spatially pooled pairs of
ensemble forecasts and corresponding nowcasts from the previous 20
days. The nowcast\footnote{Generally, the term nowcast is used for
short-term weather forecasts, comprising prediction horizons from 0 to
6 hours ahead. Here we use it for the initialization of the ECMWFs
control run---a\ distinguished NWP run outside the 50-member core
ensemble considered here---that represents the best estimate of the
state of the atmosphere at the initialization time, given recent and
concurrent observational assets. In our specific usage, the term
nowcast thus corresponds to a prediction horizon of 0 hours, and it
provides a single-valued best estimate of the state of the atmosphere,
rather than an ensemble.} that serves as grid-based ground truth is
the corresponding initialization of the ECMWFs so-called control run
\citep{Molteni1996}. The members of the unprocessed raw ECMWF
ensemble appear to capture spatial structure fairly well, but they
show an overall negative bias, especially in the mountainous Alps
region in the south and in the central east of the country. While the
BMA postprocessing addresses biases, and the use of a single BMA model
avoids inconsistencies between the univariate postprocessed predictive
distributions themselves, the independent samples result in noisy and
incoherent spatial structure. The ECC postprocessed ensemble inherits
the bias-corrected marginals from the independent BMA postprocessed
forecast and simultaneously maintains the $L = 1221$ variate
dependence structure in the raw ensemble.

While these examples concern the spatial case only, ECC is equally
well suited to handling temporal and cross-variable dependencies, with
Figure~\ref{figECC} illustrating the latter aspect. To generate
physically realistic and consistent ensemble forecasts of temporal
trajectories, constraints can be put on the BMA or NR parameters, so
that they vary smoothly across lead times, which ensures the temporal
consistency of the postprocessed marginal predictive distributions.
Then, the ECC approach can be used to account for dependence
structures across lead times. These settings are being investigated
in ongoing work, and we expect to report quantitative results in due
time.

\section{Discussion} \label{secdiscussion}

The intensified attention to the quantification of uncertainty in the
output of complex simulation models poses major challenges in a vast
range of critical applications. In this paper, we have introduced the
general uncertainty quantification framework of ensemble copula
coupling (ECC), which we have illustrated on the key example of
numerical weather prediction (NWP). The approach is conceptionally
very simple and straightforward to implement in practice. Starting
from raw ensemble output, ECC employs standard techniques to obtain
postprocessed predictive distributions for each of the univariate
margins individually. Then we quantize the postprocessed predictive
distributions and adopt the rank dependence structure of the raw
ensemble, as embodied by its empirical copula.

The defining feature of the ECC approach, namely, the adoption of the
rank order structure of the raw ensemble, also sets its limitations.
The number of members in the ECC postprocessed ensemble equals that of
the raw ensemble, which typically is small, and ECC operates under a
perfect model assumption with respect to the multivariate rank
dependence structure. For state-of-the-art NWP models such an
assumption seems defensible and reasonably adequate in practice, and
it can be comfirmed by diagnostic checks, as we have illustrated in
Figure~\ref{figmotivation}, where the situation might be typical, but
cannot be expected to be encountered each and every day. Generally,
it seems realistic to assume that numerical models may show errors in
dependence structures, which one may wish to diagnose and ameliorate
to the extent possible. Future work in these directions is strongly
encouraged.

Currently, approaches of the ECC type are being investigated and
tested by weather centers internationally; see, for example, the
recent work of \citet{Flowerdew2012}, \citet{Pinson2011} and
\citet{RoulinVannitsem2012}. We applaud these developments and call
for case studies and quantitative comparisons to the Schaake shuffle
\citep{Clark2004}, which also admits an empirical copula
interpretation. In ECC, the multivariate dependence structure of the
forecast errors derives from the ensemble forecast; in the Schaake
shuffle, it derives from a record of historical weather observations.
Judiciously designed combinations of the ECC and the Schaake shuffle
approaches address the aforementioned problem of the statistical
correction of systematic errors in dependence structures, and thus
might lead to improved predictive performance.

If the model output under consideration is low-dimensional or strongly
structured, parametric copula approaches become available, which may
allow for the correction of any systematic errors in the ensemble's
representation of conditional dependence structures. Here, the most
prominent option lies in the use of Gaussian copulas, as in the general
approach of \citeauthor{Moeller2012}\break  (\citeyear{Moeller2012}) and in the temporally or spatially
structured settings of \citet{Gel2004}, Berrocal, Raftery and Gneiting
(\citeyear{Berrocal2007,Berrocal2008}) and \citet{Pinson2009}. In such
situations, it is to be expected that parametric techniques outperform
the ECC approach and the Schaake shuffle, and comparative studies of
the predictive abilities and relative merits of the various methods are
strongly encouraged. Given its intuitive appeal and simplicity of
implementation, the ECC approach offers a natural benchmark.

In Figure~\ref{figspatialtemp} we have given an example of how ECC
can be used to restore spatial consistency in weather field forecasts
directly on the model grid. The aforementioned parametric Gaussian
approaches of \citet{Gel2004} and \citet{Berrocal2007} can achieve
this, too, but require elaborate spatial statistical models to be
fitted. In contrast, the computational and human resources
necessitated by ECC are nearly negligible, and ECC can also handle
temporal and cross-variable dependencies, for model output of nearly
any dimensionality.

While we have focused on weather forecasting in this paper, the
general framework of ECC as a multi-stage approach to the
quantification of uncertainty in the output of complex simulation
models with intricate multivariate dependence structures is likely to
be useful in a vast range of applications. Essentially, ECC can be
applied whenever an ensemble of simulation runs is available, the
ensemble is capable of realistically representing multivariate
dependence structures, and training data for the statistical
correction of the univariate margins are at hand. In this general
setting of uncertainty quantification, the goals articulated by
\citet{GneitingBalabdaouiRaftery2007} continue to provide guidance, in
that we seek to gauge our incomplete knowledge of current, past or
future quantities of interest by means of joint probability
distributions, which ought to be as sharp as possible, subject to them
being calibrated, in the broad sense of reality being statistically
compatible with the postprocessed distributions.

\section*{Acknowledgments}

We are indebted to colleagues at Heidelberg University, the University
of Washington, the German Weather Service (DWD), the European Centre
for Medium-Range Weather Forecasts (ECMWF) and elsewhere, including
but not limited to Konrad\break  Bogner, Jonathan Flowerdew, Renate Hagedorn,
Tom Hamill, Alex Lenkoski, Martin Leutbecher, Florian Pappenberger,
Pierre Pinson, David Richardson and Johanna Ziegel, who have
graciously shared their thoughts and expertise. In particular, Tom
Hamill drew our attention to approaches of the ECC type during a
stroll on the University of Washington campus in summer 2009, and
Martin Leutbecher noted a data error in a poster version of our work
presented at a conference in 2012. We gratefully acknowledge support
by the Volkswagen Foundation and sfi$^2$, Statistics for Innovation in
Oslo, and we thank the Editors and referees for their constructive
feedback.

\begin{supplement}
\stitle{Dynamic version of Figure~\ref{figECC}\\}
\slink[doi]{10.1214/13-STS443SUPP} 
\sdatatype{.pdf}
\sfilename{sts443\_supp.pdf}
\sdescription{In this version of Figure~\ref{figECC}, the
ensemble reordering step in the ECC approach is \mbox{elucidated} when
switching back and forth between pages.}
\end{supplement}

%
%

\end{document}